\documentclass[numberedappendix,apj,twocolumn]{emulateapj}

\shorttitle{IUDF10: Rest-frame Optical View on $z\sim4$ Galaxies}
\shortauthors{Oesch et al.}

\submitted{Draft version \today}

\begin{document}

\title{A Rest-frame Optical View on $z\sim4$ Galaxies I: Color and Age \\Distributions from Deep IRAC Photometry of the IUDF10 and GOODS Surveys
\altaffilmark{1}}

\altaffiltext{1}{Based on data obtained with the \textit{Hubble Space Telescope} operated by AURA, Inc. for NASA under contract NAS5-26555. Based on observations with the Spitzer Space Telescope, which is operated by the Jet Propulsion Laboratory, California Institute of Technology under NASA contract 1407.}

\author{P. A. Oesch\altaffilmark{2,\dag},
I. Labb\'{e}\altaffilmark{3}, 
R. J. Bouwens\altaffilmark{3}, 
G. D. Illingworth\altaffilmark{2}, 
V. Gonzalez\altaffilmark{2,4},
M. Franx\altaffilmark{3}, 
M. Trenti\altaffilmark{5}, \\
B. P. Holden\altaffilmark{2},
P. G. van Dokkum\altaffilmark{6},
D. Magee\altaffilmark{2}
}

\altaffiltext{2}{UCO/Lick Observatory, University of California, Santa Cruz, CA 95064; poesch@ucolick.org}
\altaffiltext{3}{Leiden Observatory, Leiden University, NL-2300 RA Leiden, Netherlands}
\altaffiltext{4}{University of California, Riverside, CA, USA}
\altaffiltext{5}{Institute of Astronomy, University of Cambridge, Madingley Road, Cambridge CB3 0HA, UK}
\altaffiltext{6}{Department of Astronomy, Yale University, New Haven, CT 06520}
\altaffiltext{\dag}{Hubble Fellow}

\begin{abstract}
We present a study of rest-frame UV-to-optical color distributions for $z\sim4$ galaxies based on the combination of deep HST/ACS+WFC3/IR data with Spitzer/IRAC imaging. In particular, we use new, ultra-deep data from the IRAC Ultradeep Field program (IUDF10). Our sample contains a total of $\sim2600$ galaxies selected as $B$-dropout Lyman Break Galaxies (LBGs) in the HUDF and one of its deep parallel fields, the HUDF09-2, as well as GOODS-North and South. This sample is used to investigate the UV continuum slopes $\beta$ and Balmer break colors ($J_{125}-[4.5]$) as a function of rest-frame optical luminosity. The [4.5] filter is chosen to avoid potential contamination by strong rest-frame optical emission lines. 
We find that galaxies at $M_{z}<-21.5$ (roughly corresponding to $L^*_{z\sim4}$) are significantly redder than their lower luminosity counterparts. 
The UV continuum slopes and the $J_{125}-[4.5]$ colors are well correlated. The most simple explanation for this correlation is that the dust reddening at these redshifts is better described by an SMC-like extinction curve, rather than the typically assumed Calzetti reddening. After correcting for dust, we find that the galaxy population shows mean stellar population ages in the range $10^{8.5}$ to $10^{9}$ yr, with a dispersion of $\sim$0.5 dex, and only weak trends as a function of luminosity. In contrast to some results from the literature, we find that only a small fraction of galaxies shows Balmer break colors which are consistent with extremely young ages, younger than 100 Myr. Under the assumption of smooth star-formation histories, this fraction is only 12-19\% for galaxies at $M_{z}<-19.75$. Our results are consistent with a gradual build-up of stars and dust in galaxies at $z>4$ with only a small fraction of stars being formed in short, intense bursts of star-formation.
\end{abstract}


\section{Introduction}

Great progress has been made over the last decade in exploring galaxies at very early times.
Large optical and NIR surveys with HST, such as the HUDF/HUDF09 \citep{Beckwith06,Oesch07,Bouwens11c}, GOODS/CANDELS \citep{Giavalisco04a,Grogin11,Koekemoer11}, or BORG \citep{Trenti11,Bradley12} allowed us to identify large numbers of galaxies at $z\sim4-8$ which are used to study the galaxy build-up in the first 1.5 Gyr of cosmic time. 
  
One of the main shortcomings of HST surveys, however, is that they only probe the rest-frame UV light of high-redshift galaxies. Even the recently installed NIR camera WFC3/IR only probes to $\lesssim 2600$ \AA\ at $z>4$ and the crucial rest-frame optical wavelengths are out of reach of HST. The observed light of high-redshift galaxies is thus strongly susceptible to dust extinction.

\begin{deluxetable*}{lcccc}
\tablecaption{Data and Sample used in this Paper\label{tab:data}}
\tablewidth{0 pt}
\tablecolumns{5}
\tablehead{\colhead{Field} & Area [arcmin$^2$] & $3\sigma$ Depth [3.6] & $3\sigma$ Depth [4.5] & Galaxy Sample  (total / $>3\sigma$ in [4.5])  }

\startdata
HUDF  & 4.7  & 27.1  & 26.8  & 230 / 59 \\ 
HUDF09-2  & 4.7  & 26.8  & 26.4  & 118 / 23 \\ 
GOODS-S  & 160  & 26.2  & 25.7  & 1635 / 813 \\ 
GOODS-N  & 120  & 26.2  & 25.7  & 660 / 378 

\enddata

\tablecomments{Depths are total AB magnitudes. }
\end{deluxetable*}

The only way to efficiently probe the rest-frame optical light of faint $z>4$ galaxies is with the $Spitzer$ Space Telescope, which samples up to $\sim9000$ \AA\ at $z>4$  in its 4.5 \micron\ channel. Several surveys have obtained deep imaging in the [3.6] and [4.5] bands over a number of fields to cover the rest-frame optical of the large samples of high-redshift galaxies that were identified with HST.  
Such data are crucial for an unbiased study of the galaxy stellar mass build-up at $z\gtrsim4$. They were already used to measure stellar population properties and galaxy masses at $z\sim4-7$ \citep[e.g.][]{Eyles07,Verma07,Stark09,Yabe09,Labbe10a,Labbe10b,Gonzalez10,Gonzalez12a,Gonzalez12b,Lee12}, and recently even out to $z\sim8$ \citep{Labbe12,Yan11}.

Despite some debate in the early literature, the rest-frame UV data alone indicate that UV bright galaxies at $z>2$, on average, show redder UV spectral slopes $\beta$ than their lower luminosity counterparts \citep[e.g.][]{Reddy08,Bouwens09b,Bouwens12a,Castellano11,Dunlop11,Finkelstein12}.
It is now clear that the galaxies at the bright end of the UV LF show significantly redder colors than the average galaxy already as early as $z\sim4-6$ \citep[e.g.][]{Lee11,Gonzalez12a,Willott12}.
This trend is likely caused by an increase in dust extinction at bright luminosities, since small amounts of dust have strong effects on the rest-frame UV colors \citep[e.g.][]{Wilkins11b,Bouwens12a}.

When including $Spitzer$/IRAC photometry to measure rest-frame UV-to-optical colors as a function of UV luminosity, \citet{Gonzalez12a} and \citet{Lee11} find that these colors are broadly consistent with being just due to dust extinction assuming a similar stellar population age at different UV luminosities. Such a measurement can be significantly biased, however, if the galaxy population shows large scatter in dust extinction at a fixed stellar mass.
An obvious next step is therefore to study galaxy colors as a function of rest-frame optical luminosity, which is significantly less affected by dust extinction and is a much more reliable proxy for galaxy stellar mass.

The main reason that such studies are not readily done is the difficulty of measuring accurate IRAC photometry for faint high-redshift sources, given the wide IRAC PSF of $\sim1\farcs6$ ($\sim10\times$ the HST PSF). Without a neighbor-subtraction scheme, only about $30-40\%$ of galaxies are isolated enough for reliable aperture photometry \citep[e.g.][]{Stark09}, resulting in small sample sizes and low number statistics. Several groups have therefore developed sophisticated techniques to obtain photometry from crowded IRAC data over the last few years \citep[see e.g.][]{Labbe06,Grazian06,Laidler07}.

Another complication of rest-frame optical studies from broad-band IRAC photometry at $z>4$ are strong emission lines. There is now ample evidence that rest-frame optical lines can provide a very significant contribution to the observed $Spitzer$/IRAC fluxes of up to 0.5 mag \citep[e.g.][]{Schaerer09,Shim11,Stark12,Gonzalez12a,Labbe12}. Such lines have to be taken into account when interpreting galaxy colors or when fitting spectral energy distributions (SEDs). 
For instance, using synthetic models with emission lines \citet{Schaerer09} and \citet{deBarros12} estimate that the majority of galaxies at $z\sim4-7$ are best fit with extremely young galaxy ages\footnote{If not mentioned otherwise, throughout this paper, we define age as the time since the onset of star-formation.} of $<100$ Myr. This would have very important consequences on the total inferred cosmic star-formation rate density. However, it is a concern whether the many high-quality photometric constraints in the rest-frame UV compared to the few data points in the rest-frame optical could be biasing the ages determined from standard SED fits. 

A simple way to test such predictions is to study galaxy rest-frame UV-to-optical color distributions. 
In this paper, we use deep $Spitzer$/IRAC imaging for a detailed investigation of galaxy color distributions of a large sample of $z\sim4$ galaxies as a function of rest-frame optical luminosities. With these data we study the origin of the UV continuum reddening of galaxies, and we constrain the fraction of galaxies that are consistent with extremely young ages ($<100$ Myr) based on dust-corrected UV-to-optical colors.

Throughout this paper, we adopt $\Omega_M=0.3, \Omega_\Lambda=0.7, H_0=70$ kms$^{-1}$Mpc$^{-1}$, i.e. $h=0.7$. Magnitudes are given in the AB system \citep{Oke83}.

\section{Data and Galaxy Sample}
\label{sec:data}

\subsection{Observations}

Here we analyze $HST$ and $Spitzer$ data over four different fields: the two wide-area GOODS/CANDELS fields, the ultra-deep HUDF, and its deep parallel, the HUDF09-2\footnote{The second HUDF09 parallel field to the North-East is not used due to its lack of $B_{435}$ imaging, which is needed for the $z\sim4$ LBG selection.}. 
All these fields are covered by ACS imaging from $B_{435}$ to $z_{850}$ from the UDF05 \citep[PI: Stiavelli][]{Oesch07,Oesch09}, the HUDF \citep{Beckwith06} and GOODS surveys \citep{Giavalisco04b}. WFC3/IR data in three filters including $J_{125}$ and $H_{160}$, was obtained as part of the HUDF09 (PI: Illingworth; Bouwens et al. 2011), the ERS (PI: O'Conell; Windhorst et al. 2011), and the CANDELS MCT programs (PI: Faber/Ferguson; Grogin et al. 2012, Koekemoer et al. 2012). For GOODS-North, we only include the first six visits obtained until July 19, 2012. The data reach to limits of $H_{160} = 27.0 - 28.5$ mag over the $\sim300$ arcmin$^2$ GOODS fields, and $H_{160} = 29.3-29.9$ mag over the two HUDF09 fields (covering a total of $\sim10$ arcmin$^2$).

Most importantly, all fields are covered by deep $Spitzer$/IRAC \citep{Fazio04} imaging in the [3.6] and [4.5] channels. In particular, we use new ultra-deep data from the 262 h IRAC Ultradeep Field program (IUDF10; PI: Labb\'{e}; see also Labb\'e et al. 2012). This survey pushes the IRAC [3.6] and [4.5] imaging to exposure times of $\sim120$ h and $\sim80-120$ h over the HUDF and the HUDF09-2, respectively. The two wider-area GOODS fields were previously covered by 23 h of public IRAC imaging (M. Dickinson et al. in prep).

All the data and samples used in this paper are summarized in Table \ref{tab:data}.

\subsection{Measurements of the $z\sim4$ Galaxy Sample}

The galaxy sample analyzed here is identified using a standard $z\sim4$ Lyman Break Galaxy (LBG) $B_{435}$-dropout selection based on the HST ACS data \citep[see, e.g.,][]{Giavalisco04a,Bouwens07}. Sources were identified in a $\chi^2$ image based on the $V_{606}$, $i_{775}$, and $z_{850}$ ACS data, before applying the following color selection criteria:
\begin{eqnarray*}
(B_{435} - V_{606}>1.1)& \wedge& B_{435} - V_{606} > (V_{606} - z_{850}) + 1.1 \\
 & \wedge & (V_{606} - z_{850} <1.6) 
\end{eqnarray*}

This selects galaxies at $z\sim3.5-4.5$, with a mean redshift $\bar{z}=3.8$ and results in a total sample of 2643 galaxies down to $M_{UV} = -16$ mag over all four fields. HST photometry is obtained in small elliptical Kron apertures from PSF matched images \citep[see][for details]{Bouwens12a}. The IRAC photometry is derived following the procedure of \citet{Labbe10b,Labbe10a}. In particular, we subtract neighboring foreground sources based on their profiles in the HST $J_{125}$ and $H_{160}$ images, after convolution to the IRAC PSF. We then perform aperture photometry on the cleaned images in 2\arcsec\ diameter apertures, and correct to total fluxes using the growth curves of nearby stars in the field. The typical corrections are a factor 2.2, consistent with expectations from the IRAC photometry handbook. A comparison of our photometry against the GOODSMUSIC catalog over GOODS-South \citep{Santini09} shows no bias and a scatter of $0.20-0.25$ mag. 

The automated IRAC cleaning procedure does not always work. In particular, if a source is too close to an extremely IRAC-bright neighbor, reliable photometry can not be extracted. We therefore inspect all sources by eye for failures in the neighbor subtraction and flag those. These are excluded from the final analysis. With our procedure we are able to obtain clean IRAC photometry for $\sim75\%$ of all sources. This is compared to a $\sim30-40\%$ success rate for simple aperture photometry on isolated sources only. We therefore more than double the galaxy sample compared to previous analyses which omit such a neighbor subtraction.
Details on the IRAC cleaning and photometry will be presented in Labb\'{e} et al. (in preparation).

Since the IRAC flux in [3.6] can be significantly affected by strong H$\alpha$ emission for a large fraction of our $z\sim3.5-4.5$ sample \citep[e.g.][]{Shim11,Stark12}, we restrict our analysis to IRAC [4.5] band detections. This samples rest-frame 9000 \AA, which is devoid of strong line emission. Our galaxy sample consists of 1273 sources with IRAC [4.5] detections more significant than $3\sigma$. 82 of these come from the ultra-deep IUDF10 data. 

Using the observed [4.5] fluxes, we compute the rest-frame absolute magnitude at 9000 \AA, assuming a fixed redshift for all sources at the mean of our sample $\bar{z}=3.8$. In the following, we will denote this as $M_z$.

One of the main goals of our analysis is to identify the origin of the UV continuum reddening in these galaxies. To this end, we compute the UV continuum slopes, $\beta$ ($f_\lambda \propto \lambda^\beta$), as in \citet{Bouwens12a} from a power-law fit to the observed fluxes in the HST bands $V_{606}, ~ i_{775}, ~z_{850},~Y_{098/105}$ (where available), and  $J_{125}$.

\begin{figure}[tbp]
	\centering	
	\includegraphics[width=\linewidth]{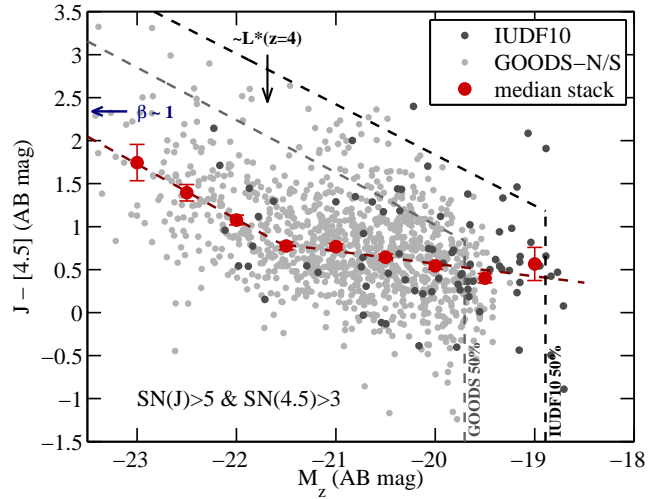}
  \caption{The observed Balmer break color $J_{125} - [4.5]$ (sampling rest-frame 2500~\AA$ - $9000~\AA) vs. rest-frame optical magnitude for $z\sim4$ galaxies. The gray dots represent galaxies with $>3\sigma$ detections in $[4.5]$ and $>5\sigma$ in $J_{125}$ (dark gray: IUDF10, light gray: GOODS). The median colors in bins of $M_z$ are indicated by red dots with errorbars computed from bootstrap resampling.  A clear luminosity dependence is evident, with a significant steepening at the bright end, where the Balmer break colors become increasingly redder. The increase in reddening sets in at around $M_{z}=-21.5$, which roughly corresponds to $L^*_z(z=4)$, as estimated based on the characteristic magnitude of the UV LF and on the average UV-to-optical colors. This is indicated by the vertical black arrow. 
The dashed red line represents two linear fits to the $J_{125} - [4.5]$ vs $M_z$ relation at $M_{z}<-21.5$ and $M_{z}>-21.5$.  
  Shown as gray and black dashed lines are the expected 50\% completeness limits of the GOODS and IUDF10 datasets, respectively. The horizontal blue arrow shows the approximate location of galaxies with UV continuum slopes $\beta=1$ according to the relation we find later in Figure \ref{fig:betaJ1}. The LBG selection volume is significantly reduced for sources redder than that.
  }
	\label{fig:fig2}
\end{figure}

\section{Results}

\subsection{Galaxy Colors as a Function of Rest-Frame Optical Luminosity}

Our large sample of galaxies with robust IRAC photometry allows us to investigate galaxy colors as a function of rest-frame optical luminosity. 
Of particular interest is the $J_{125}-[4.5]$ color, which straddles the Balmer break. At $z\sim4$, $J_{125}$ samples 2500 \AA, and [4.5] samples 9000 \AA. Note that $J_{125}$ is chosen over $H_{160}$ because the latter is contaminated by light longward of 4000 \AA\ for the lower redshift tail of the B-dropout distribution, which complicates its interpretation.

The $J_{125}-[4.5]$ color-magnitude diagram is shown in Figure \ref{fig:fig2}. This plot reveals that more luminous galaxies are significantly redder than their lower luminosity counterparts. Galaxies at $M_z = -22.5$ show colors of $J_{125}-[4.5] = 1.4$, while galaxies at $M_z \sim -20$ have $J_{125}-[4.5] = 0.6$. In particular, at $M_z>-21.5$ the color distribution is well-described by a Gaussian with an observed dispersion of $0.4-0.5$ mag (see also later Figure \ref{fig:ColHist}).

Interestingly, a luminosity dependence in the color-magnitude diagram is seen at all luminosities. However, the dependence is significantly steeper at the brightest luminosities. The color-magnitude relation is well reproduced by a two component linear relation, with a break roughly at $L_*(z=4)$\footnote{$L*$ is crudely estimated based on the average $i_{775}-[4.5]$ color and the characteristic luminosity of the UV LF at $z=4$ of $M^*_{UV} = -21.06$ \citep{Bouwens07}. This results in an estimated $M^*_z = -21.7$ mag, which is significantly fainter than the characteristic luminosity of the rest-frame $V$-band LF as measured by \citet{Marchesini12} who find $M^*_V = -22.76^{+0.40}_{-0.63}$ mag.}.  The slope of the relation changes from $-0.64\pm0.02$ at $L\gtrsim L*$ to $-0.15\pm0.05$ to $L\lesssim L*$.

\begin{figure}[tbp]
	\centering
	\includegraphics[width=\linewidth]{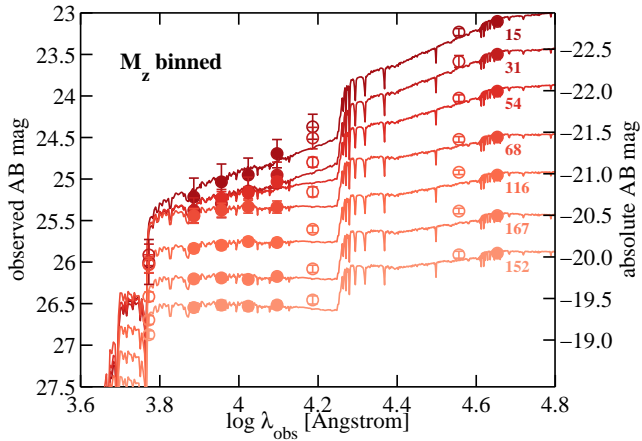}
  \caption{Median stacked SEDs of galaxies in bins of $z$-band luminosities. The numbers indicate how many galaxies contribute to each stack. Dramatic trends toward redder colors are seen as a function of rest-frame optical luminosities, in particular at $M_z<-21.5$. Galaxies are clearly not self-similar as is inferred from studies based on binning galaxies in UV luminosities only \citep[e.g.][]{Gonzalez12a}. The open symbols indicate filter fluxes which are not used when deriving synthetic SED fits to the stacks, since they are expected to change significantly across the extent of the LBG redshift selection. The $H_{160}$-band probes longward of 4000 \AA\ for the lower redshift tail of the selection, while strong H$\alpha$ emission may contaminate the [3.6] measurement for 50\% of the sample (at $z=3.8-5.0$). As can be seen, both these filter fluxes are indeed significantly higher than the best-fit SED for all stacks.
  }
	\label{fig:stackedSED}
\end{figure}

Previous studies that were based on binning galaxies as a function of rest-frame UV luminosities suggested that the average UV-to-optical SED of galaxies changes only very little with luminosity. For instance, \citet[][]{Gonzalez12a} find rest-frame $U-V$ colors that get redder by only 0.026 per mag in UV luminosity.
Such weak trends were in good agreement with the suggestion from SPH simulations that found star-formation histories of galaxies to be essentially self-similar \citep{Finlator11a}. 

Our findings above suggest that using rest-frame optical luminosities to bin and stack galaxy SEDs, much more significant trends should become apparent. This is indeed the case as shown in Figure \ref{fig:stackedSED}, where we present SEDs in bins of $M_{z}$. A significant increase in reddening toward brighter $z$-band luminosities is obvious in these stacks. While these SEDs span 3 mag in $M_z$, they only vary by $\sim1.5$ mag in UV luminosity. 
Furthermore, the stacked SEDs show a clear correlation between the UV continuum slope and the UV-to-optical color.  
This indicates that galaxies that are luminous in the rest-frame optical are significantly more dust 
obscured than fainter systems, which is likely the origin for the steeper trend in the $J_{125}-[4.5]$ color we found in Figure \ref{fig:fig2}.

Finally, from these stacks it is clear that H$\alpha$ flux contamination is indeed a potential problem for our $B$-dropout sample. All stacked SEDs show a significant excess at 3.6~$\mu$m compared to the best-fit SED, ranging from 0.06 mag to 0.14 mag, with a tentative trend to a larger excess at fainter luminosities. Since H$\alpha$ is only present in the [3.6] filter at $z=3.8-5.0$ (i.e. in $\sim50\%$ of our redshift selection), these numbers are roughly consistent with the mean excess of 0.27 mag as inferred by \citet{Stark12}, in particular for fainter sources.

\begin{figure}[tbp]
	\centering
	\includegraphics[width=\linewidth]{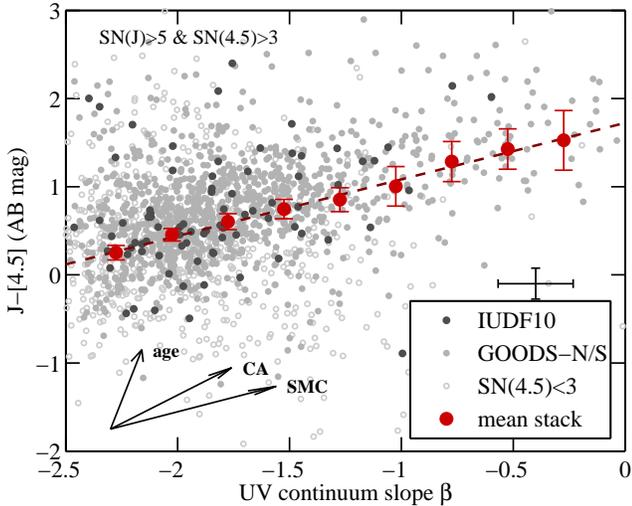}
  \caption{The relation between the UV continuum slope $\beta$ and the observed $J-[4.5]$ color for $z\sim4$ galaxies. As in Figure \ref{fig:fig2}, the filled dots correspond to sources with $>5\sigma$ detections in $J_{125}$ and S/N($[4.5]$)$>3$. Dark and light grey dots denote galaxies in the IUDF10 and GOODS-North/South respectively. Open circles correspond to lower significance IRAC fluxes ($<3\sigma$) in either sample.
  The mean stacked color (including the measured fluxes of IRAC undetected galaxies) in bins of $\beta$ are indicated as red dots with errorbars corresponding to the uncertainty on the mean.
The red dashed line indicates a linear fit to these stacked colors. 
   The correlation between $\beta$ and the $J-[4.5]$ color is primarily driven by dust extinction, as shown by the vectors in the lower left. CA indicates the Calzetti et al. 2000 star-burst dust reddening vector with $A_V=0.5$ mag, and SMC represents the Pei 1992 reddening vector with $A_V=0.25$ mag. Also shown is an age vector which is derived for an SED with constant star-formation and an age from 0.1 to 1 Gyr. 
  }
	\label{fig:betaJ1}
\end{figure}

A potential problem with the above measurements are the combination of different selection effects that can result in biased distributions. In Figure \ref{fig:fig2}, we show the estimated completeness limits of our data as dashed lines.
The vertical limit is due to the IRAC S/N cut, while the slanted line is mostly caused by the $J_{125}$-band cut, but its slope is influenced by the correlation between $J_{125}-[4.5]$ vs $\beta$ (see Figure \ref{fig:betaJ1}) due to the LBG selection. The completeness lines are estimated based on simulating a large number ($10^5$) of galaxies with synthetic SEDs with $M_z = -24$ to $-18$, with UV slopes in the range $\beta=-3$ to $1$, and with redshifts according to our redshift selection function \citep[see e.g.][]{Bouwens07}. The fluxes of these SEDs are then perturbed with Gaussian dispersions and all the selection criteria are applied, from which we compute the completeness as a function of $M_z$ and $J_{125}-[4.5]$. 

Clearly, at faint luminosities, we are potentially biased against red sources in the GOODS data. However, the $J_{125}-[4.5]$ color distribution peaks at sufficiently bluer colors than the completeness limit. Using the simulations above, we estimate that the observed mean color at $M_z=-20$ is not significantly biased due to selection effects ($\lesssim0.05$ mag).

Finally, photometric errors in [4.5] might bias the measurement of the color-magnitude relation of Figure \ref{fig:fig2}, since both axes depend on the [4.5] flux measurement. A positive error in [4.5] would result in a brighter $M_z$ with a redder $J_{125}-[4.5]$ color. We therefore checked and confirmed that an analogous steepening in the color-magnitude relation is still seen if the absolute magnitude derived from the [3.6] photometry is used instead of $M_z$. This relation is not shown here due to the possible flux contamination from strong H$\alpha$ lines, as explained earlier. Nevertheless, this test suggests that the steepening in the color-magnitude relation is not significantly affected by scatter in the [4.5] flux measurements.

\begin{figure}[tbp]
	\centering
	\includegraphics[width=\linewidth]{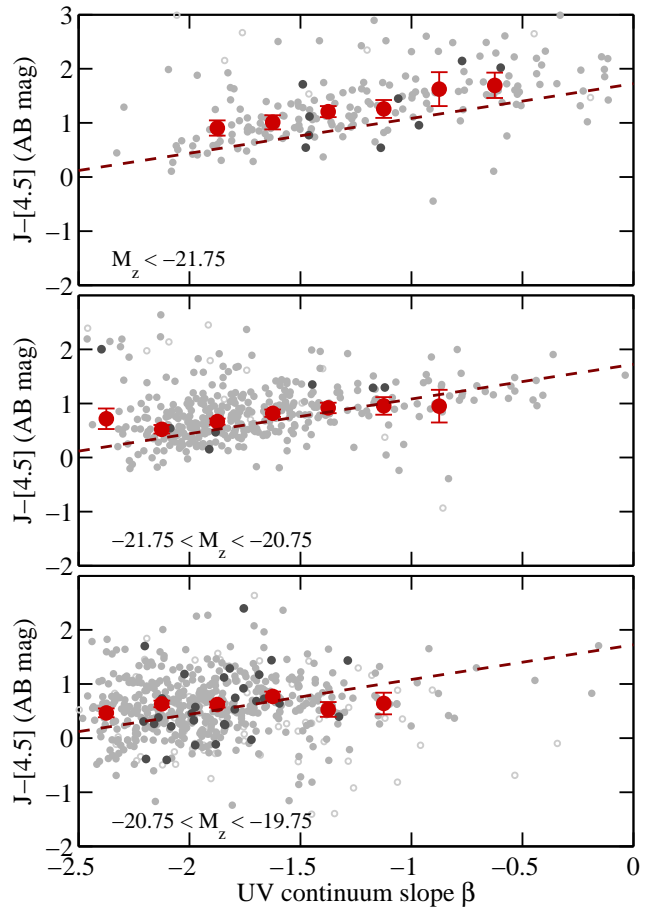}
 \caption{The relation between the UV continuum slope $\beta$ and the observed $J-[4.5]$ color for $z\sim4$ galaxies as a function of 9000 \AA\ luminosity. The dashed red line is the mean relation derived for all galaxies (Equation \ref{eq:1}) and is repeated in every panel.
As in Figure \ref{fig:betaJ1}, the filled dots correspond to sources with $>5\sigma$ detections in $J_{125}$ and S/N($[4.5]$)$>3$. Dark and light grey dots denote galaxies in the IUDF10 and GOODS-North/South respectively. Open circles correspond to lower significance IRAC fluxes ($<3\sigma$) in either sample. 
The mean stacked color (including the measured fluxes of IRAC undetected galaxies) in bins of $\beta$ are indicated as red dots with errorbars corresponding to the uncertainties on the means. Only bins with more than 15 sources to stack are shown.
  }
	\label{fig:betaJ1_fL}
\end{figure}

\subsection{UV Continuum Slopes and Balmer Breaks}
\label{sec:UVslopesBalmerBreaks}

In order to investigate the origin of red UV continuum slopes $\beta$ in brighter galaxies, we study the relation between $\beta$ and the $J_{125} - [4.5]$ color of $z\sim4$ galaxies, which effectively compares the reddening in the UV with the Balmer break. 

As shown in Figure \ref{fig:betaJ1}, there is a clear and relatively tight relation between the UV slope $\beta$ and the $J_{125}-[4.5]$ color, with UV redder galaxies showing redder colors across their Balmer break. The relation is well-fit by the linear relation: 
\begin{equation}  
J_{125} - [4.5] = (0.64\pm0.03) \times \beta + (1.72\pm0.04)
\label{eq:1}\end{equation}
where we have included all galaxies with $J_{125}$ detections $>5\sigma$, irrespective of their [4.5] detection significance. Excluding sources with non-detections in the IRAC [4.5] band results in a relation with an identical slope, but with a slightly redder zeropoint (1.86$\pm0.04$). 

For sources with $\beta<-1.5$, the color dispersion perpendicular to the dust reddening is only $0.4$ mag, which increases to $0.5-0.6$ mag for redder sources. These dispersions are most likely explained by a variation in stellar population ages as indicated by the age vector in Figure \ref{fig:betaJ1} (see also Section \ref{sec:youngfrac}).

Interestingly, a \citet{Calzetti00} dust curve would lead to a steeper relation than what is seen in the data. This can be seen from the dust reddening vector shown in the lower left of Figure \ref{fig:betaJ1}. Calzetti dust results in a slope of 1.29 and is clearly inconsistent with the relation given above. This means that the correlation between $\beta$ and $J_{125}-[4.5]$ can not simply be explained by Calzetti dust reddening only, but would require that dustier galaxies show younger stellar population ages in a correlated manner which results exactly in the observed slope.

A simpler explanation of the $J_{125}-[4.5]$ vs $\beta$ correlation, however, is that galaxies are instead reddened by SMC-like dust. Using the parametrization of \citet[][]{Pei92} SMC reddening results in a slope of 0.66, which is in excellent agreement with the observed relation (see also the reddening vector in Figure \ref{fig:betaJ1}). Our data therefore provides tentative evidence that SMC reddening might be more appropriate for galaxies at $z\sim4$. At least for brighter galaxies, this is in agreement with recent measurements of the infrared-excess for $z\sim4$ sources with Herschel \citep{Lee12b}. At fainter luminosities, the Herschel results are consistent with the standard \citet{Meurer99} and \citet{Calzetti00} extinction instead.


We therefore test whether we find a luminosity dependence in the $J_{125}-[4.5]$ vs $\beta$ relation in our data. In Figure \ref{fig:betaJ1_fL}, we show this relation for three bins of $z$-band luminosity, down to $M_z<-19.75$, which is the completeness limit of the GOODS IRAC data.


As is evident, the brightest galaxies with $M_z<-21.75$ show a very large range of UV continuum slopes from $\beta\sim-2$ to $\beta\sim0$. Across this full range, their $J_{125}-[4.5]$ colors follow the observed slope for the full sample extremely well, consistent with the Herschel results and with a reddening of SMC-like dust. However, the colors of these bright sources are systematically shifted by about 0.3 mag to the red, likely due to systematically older stellar population ages compared to the average galaxy in our sample. The fact that there are almost no bright galaxies with $\beta<-2$ indicates that the most luminous (and thus the most massive) galaxies at $z\sim4$ must have formed both their stars and their dust over an extended period of time. 

As we move to lower luminosities, we find that the mean UV slope clearly shifts to bluer, more dust-free colors. Interestingly, in the faintest bin ($M_z\sim-20.25$) the $J_{125}-[4.5]$ color appears to be independent of $\beta$. However, this is potentially due to incompleteness, which starts affecting galaxies as blue as $J_{125}-[4.5]=0.9$ in that bin (see Figure \ref{fig:fig2}).



\subsection{Dust-Corrected UV-to-Optical Colors}
\label{sec:colorhist}

The goal of our study is to constrain the intrinsic distribution of Balmer break colors in the $z\sim4$ LBG population. Having found the correlation in Figure \ref{fig:betaJ1}, we have a means of empirically correcting the $J_{125}-[4.5]$ colors for dust reddening. Our fiducial dust correction procedure is as follows: Starting from the observed $J_{125}-[4.5]$ and $\beta$ values for galaxies, we find the intersection of the reddening vector with the track of a synthetic SED model with constant star-formation and with ages in the range $10^{7} - 2\times10^{9}$ yr. For the reddening vector, we use the empirical slope of Equation \ref{eq:1}, which is essentially equivalent to SMC reddening. 

This procedure is chosen to more accurately represent the likely dust extinction in young galaxies, which typically have bluer UV continuum slopes than the value of $-2.23$ for which the \citet[M99;][]{Meurer99} correction results in zero dust extinction.
The intrinsic, dust-corrected $\beta$ values of our fiducial approach lie in the range $-2.8$ to $-2.3$. In the appendix we show a schematic figure which explains our dust correction (Figure \ref{fig:schematicDustCorr}), and we show the effect of using the standard M99 dust correction for comparison.

\begin{figure}[tbp]
	\centering
	\includegraphics[width=\linewidth]{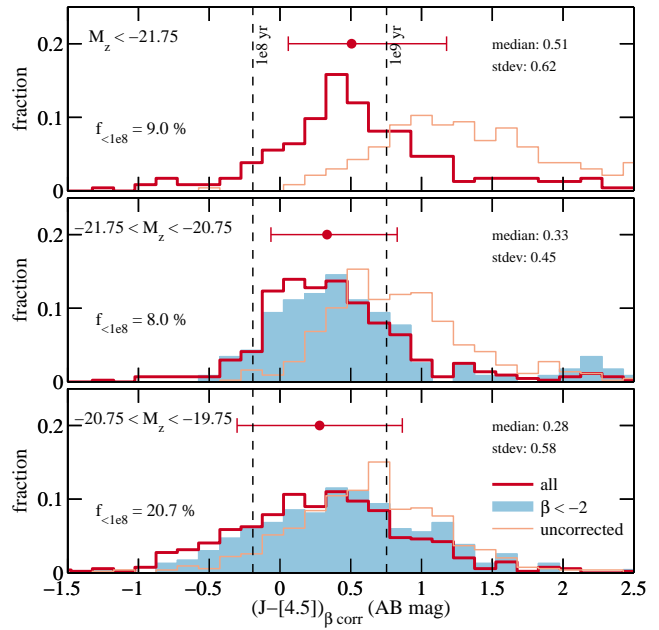}
  \caption{Histograms of dust-corrected colors $(J_{125}-[4.5])_{\beta corr}$ in three bins of rest-frame optical absolute magnitude at 9000 \AA\ (thick red line). The underlying, shaded blue histograms represent only the blue galaxy population with UV-continuum slopes $\beta<-2$ for which the dust correction is small. Also shown as thin, orange lines are the un-corrected color histograms for the full population, indicating that the dust-correction is much more significant for brighter galaxies. The dots with horizontal errorbars represent the  median and $\sim1\sigma$ range of the distribution. These indicate that unreddened galaxy colors show weak trends with rest-frame optical luminosity, but that the width of the distribution does not change dramatically. Vertical dashed lines indicate the colors of SEDs with constant star-formation rates and stellar population ages of 0.1 and 1 Gyr, which enclose the majority of the observed colors. The fraction of galaxies that are consistent with very young ages ($<0.1$ Gyr) is indicated on the left in each panel. This amounts to only $12\%$ when corrected for photometric scatter and averaged over all galaxies with $M_z<-19.75$ mag. Note, however, that the exact number depends on the assumed star-formation history.
  }
	\label{fig:ColHist}
\end{figure}

Histograms of both corrected and observed colors are shown in Figure \ref{fig:ColHist}, again for three bins of $z$-band luminosity down to $M_z=-19.75$. 
Our fiducial dust-correction results in average offsets of 0.8 mag at $M_z<-21.75$, and only 0.25 mag at $M_z = -20.25$.  These offsets in $J_{125}-[4.5]$ color correspond to $A_V = 0.41$ mag and 0.13 mag for the \citet{Pei92} SMC extinction law, respectively.

While the original colors were strongly luminosity dependent, the corrected $J_{125}-[4.5]$ colors show only a weak trend with luminosity.
The brightest sources ($M_z<-21.75$) are only $\sim0.2$ mag redder on average than galaxies at $M_z>-21.75$. Note that these histograms are well-sampled. From bright to faint, the three luminosity bins in Figure \ref{fig:ColHist} contain 234, 438, and 545 galaxies, respectively.

For the two lower luminosity bins, we also show the color histograms for blue sources with $\beta<-2$, for which the dust corrections are smallest. This provides a good consistency check. Overall, sources with $\beta<-2$ comprise 30\% of the sample at $M_z<-19.75$ (326 galaxies). However, at $M_z<-21.75$ this $\beta<-2$ sample contains only 16 sources, preventing a reliable comparison (which is why these are omitted from Figure \ref{fig:betaJ1}). In both fainter bins, the $\beta<-2$ sample shows de-reddened color distributions in excellent agreement with the full sample, indicating that our dust-correction procedure is reliable and unbiased.

\begin{figure}[tbp]
	\centering
	\includegraphics[width=\linewidth]{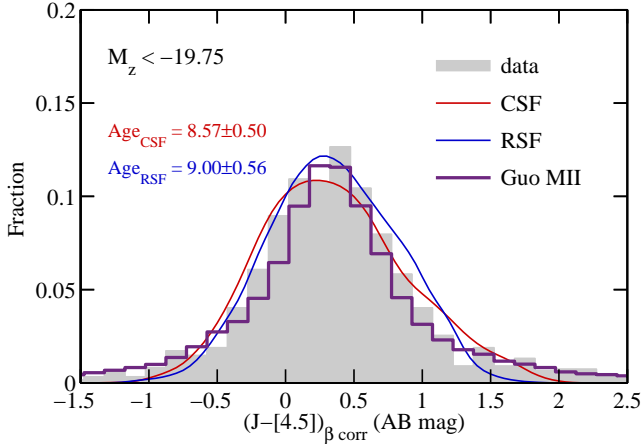}
  \caption{The color distribution for all galaxies with $M_z<-19.75$, together with three different model distributions. The filled histograms show the dust-corrected $J_{125}-[4.5]$ colors from the previous figure. The two thin lines show the best-fit color distributions when assuming a Gaussian distribution in $\log$(age) and two different SFHs (CSF: constant star-formation, RSF: a power law rising SFR $t^{1.7}$). Both reproduce the color distributions fairly well. However, they fall somewhat short at the bluest and reddest tails of the distribution. The colors of the  \citet{Guo11} SAM are shown as thick purple line. These result in a color distribution in excellent agreement with the observed corrected colors, indicating that the SAMs predict SFHs that represent reality very well.
  }
	\label{fig:ageSimHist}
\end{figure}

\subsection{Limits on the Predominance of Extremely Young Galaxy Ages}
\label{sec:youngfrac}

Several papers have recently claimed that a dominant fraction of $z>4$ galaxies may exhibit very young stellar population ages \citep[$<100$ Myr; e.g.][]{Schaerer10,deBarros12}. This would have very important consequences, since it would imply extremely short duty cycles for LBGs, and large corrections to the cosmic star-formation rate density \citep[e.g.][]{Jaacks12b}. One major difficulty in inferring stellar population ages from SED fitting is the degeneracy between dust extinction,  age, and metallicity in reproducing the reddening of the UV slope. These problems are exacerbated at $z>4$, since HST only probes rest-frame UV wavelengths, and the only rest-frame optical information typically comes from only two $Spitzer$ bands. 

Given these complications, it is thus instructive to use assumptions as simple and model independent as possible to test what range of ages is consistent with the data. This can be done using our simple dust-corrected Balmer break colors $(J_{125}-[4.5])_{\beta corr}$. In Figure \ref{fig:ColHist}, we indicate the colors of SEDs with constant star-formation and ages $10^{8}$ yr and $10^{9}$ yr. These are based on \citet{Bruzual03} models, with Salpeter IMF and sub-solar metallicities ($Z=0.2Z_\odot$). As can be seen, these lines nicely bracket the peak of the color distribution, indicating that the average $z\sim4$ LBG shows stellar population ages in this range. 

The fraction of sources with colors consistent with very young ages below $10^{8}$ yr (i.e. with $(J_{125}-[4.5])_{\beta corr}<-0.19$) is very small; less than 10\% at $M_z<-20.75$ and only 21\% at $M_z = -20.25$. Integrated over all galaxies with $M_z<-19.75$ for which we are complete, the number of galaxies with $(J_{125}-[4.5])_{\beta corr}<-0.19$ is $14$\%. 
A simple correction of this number for photometric scatter suggests that this fraction intrinsically is even smaller, i.e. 12\%. The formal uncertainty on this number is only 1\%. However, we note that this fraction is highly model dependent. In particular, using an SED model with rising star-formation rates, the young fraction is significantly reduced, due to bluer galaxy colors at a given age. Additionally, in the appendix we estimate that using the standard M99 and \citet{Calzetti00} dust relations and SEDs with constant star-formation would result in a young fraction of 19\%. Despite of these uncertainties, the observed color distributions indicate that the galaxy population may only contain a relatively small number of extremely young sources, in contrast to earlier findings in the literature based on full SED fitting.

\subsection{Fits to the Color Distribution at $z\sim4$}
\label{sec:fitstodistribution}

Using our corrected color distribution, we can set interesting constraints on the ages and star-formation histories (SFHs) of $z\sim4$ galaxies. In particular, we can fit for the distribution of galaxy ages under the assumption of smooth SFHs. This is done as follows:
We sample galaxy ages from a normal distribution in log(age) with mean $\mu_A$ and width $\sigma_A$. We then compute the expected galaxy $J_{125}-[4.5]$ color for the given age, assuming either a constant star-formation rate (CSF) or a power-law rising star-formation history ($t^{1.7}$, RSF). The latter is the population average SFH consistent with the evolution of the UV LF \citep[][]{Papovich11}. The photometry is then perturbed using the real flux uncertainties of our galaxy sample, and the resulting histogram is compared to the data using a Poissonian likelihood of the number counts. 

The best-fit color distributions are shown in Figure \ref{fig:ageSimHist}. These correspond to Gaussian age distributions with $\log(age)=8.57$ and $\sigma_{\log age} = 0.50$ for constant star-formation and $\log(age)=9.00$ and $\sigma_{\log age} = 0.56$ for the rising star-formation histories. The somewhat older mean stellar population age for the latter SFH is a result of bluer galaxy colors at a fixed stellar population age, given the dominance of younger stars.

Again, for both these distributions, we find a very small number of galaxies with ages younger than 100 Myr (only 13\% for CSF and 4\% for RSF). Interestingly, however, these distributions predict that a significant fraction of galaxies (50\% for CSF and 24\% for RSF) would not have been forming stars for longer than 400 Myr, which is the time between $z\sim5$ and $z\sim4$. These sources would therefore not have been present in the higher redshift LBG sample. The mean formation redshift of our sample is inferred to be $z_f = 5.0$ or $z_f=9.3$ assuming CSF or RSF, respectively.

While the Gaussian age distributions for both assumptions on the SFHs do reproduce the observed color histograms rather well, they fall somewhat short at the bluest and reddest tails of the distribution. This indicates that real star-formation histories contain a modest component of bursty star-formation.

\begin{figure}[tbp]
	\centering
	\includegraphics[width=\linewidth]{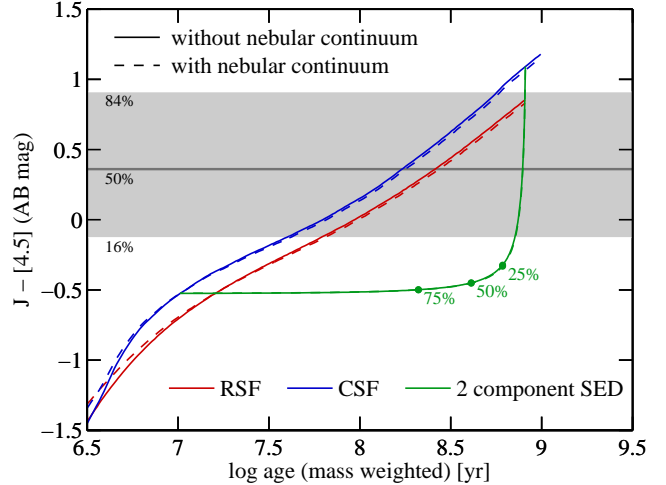}
 \caption{The $J_{125}-[4.5]$ color evolution of different SFHs as a function of mass-weighted age. The star-formation histories are: a power-law $\propto t^{1.7}$ (RSF), constant star-formation (CSF), and a two-component SFH with a varying mass-fraction of an extremely young stellar population (of age 20 Myr) on top of an older population (of age 1.6 Gyr). The location of 25\%, 50\%, and 75\% young mass fractions are indicated by small dots. Solid lines are for pure stellar continuum, while dashed lines include the UV reddening due to nebular continuum emission from hydrogen and helium recombination in the inter-stellar medium of a galaxy (assuming ionization equilibrium and an escape fraction of 50\% for ionizing photons). The shaded gray area spans 68\% ($\sim1\sigma$) of the observed galaxy color distribution at $M_z<-19.75$ mag (see Figure \ref{fig:ColHist}). The majority of galaxies clearly has less than a 25\% contribution from extremely recent bursts, and sources are consistent with a mass weighted stellar population age between 50 and 500 Myr.
  }
	\label{fig:Jm45ageSEDlibs}
\end{figure}

We can obtain a first constraint on the contribution of extremely short bursts to the total star-formation of galaxies from the expected $J_{125}-[4.5]$ color for a two-component SED with a very young population added to an older underlying one. This is shown in Figure \ref{fig:Jm45ageSEDlibs}. 
The Balmer break color saturates very quickly close to the color of the young component, even for a small mass fraction ($<10\%$) produced in a recent burst. 
Bursty star-formation would therefore significantly tilt the color distribution to young galaxy ages, even if only a small mass fraction is built in bursts. The exact numbers depend on the detailed assumptions on the underlying old population and burst ages, but this  suggests that  {\it the majority of galaxies is consistent with only a very small fraction of stars being formed in very recent bursts. }

A further constraint on bursty star-formation can be obtained from the equivalent width distribution of strong rest-frame optical lines. Based on galaxies with spectroscopic redshifts in the range $z=3.8-5.0$, \citet{Stark12} estimated the combined H$\alpha$+[SII]+[NII] equivalent widths from the IRAC [3.6] flux excess. They find that this distribution peaks at $\log(W/$\AA$) = 2.57$ and has a dispersion of 0.25 dex.
We can estimate the predicted equivalent widths from our age distribution using a simple model for hydrogen recombination. Assuming a 10\% escape fraction of ionizing photons \citep[consistent with the most recent estimates for LBGs from][]{Nestor12} and photoionization equilibrium, we can derive the H$\alpha$ emission line strengths from the tables in \citet{Osterbrock06} and from the total number of ionizing photons that are produced in the SED model. This calculation shows that our age distributions result in equivalent widths that are lower than the observed values of \citet{Stark12} by only $\sim$0.15 dex, indicating that such smooth SFHs can produce most, but not all of the rest-frame optical line strengths. Bursty star-formation is thus certainly present to produce the most extreme tail of H$\alpha$ emission, but it is likely not dominant.

Furthermore, we can use these calculations to check what fraction of the equivalent width distribution measured by \citet{Stark12} requires young galaxy ages.
For an SED with an age of 100 Myr, we derive a combined H$\alpha$+[SII]+[NII] equivalent width of 480 \AA\ and 790 \AA\ for our CSF and RSF SFH models, respectively. The \citet{Stark09} distribution contains $\sim30\%$ and $\sim10\%$ of galaxies with larger values. This provides further evidence for relatively small fractions of very young galaxies in LBG samples.

In order to compare our observations with more realistic SFHs, we use the results of the \citet{Guo11} semi-analytical model (SAM) based on the Millennium-II dark matter simulation. In order to extract the $J_{125}-[4.5]$ colors of the SAM, we fit SEDs to the dust-free absolute magnitudes of their simulated galaxies at $z=3.8$ with \citet{Bruzual03} templates, and extract the $J_{125}$ and $[4.5]$ magnitudes of the best-fit SEDs.
These intrinsic colors are then perturbed by the flux errors from our observational data. The result of this is shown as thick purple line in Figure \ref{fig:ageSimHist}. Obviously, the predicted color histogram is in very good agreement with the observed one, indicating that this SAM predicts SFHs that are a very good match to real galaxies. Note, however, that the \citet{Guo11} model does not do as well in reproducing the dust-reddened (i.e. directly observed) color distributions. Furthermore, their model falls short of bright galaxies in the LFs at blue wavelengths at $z>2$. As argued in, e.g., \citet{Henriques12} both these effects are likely connected to uncertainties in the dust modeling. Analyses similar to the ones in our paper will help to further test and refine such models.

\begin{figure}[tbp]
	\centering
	\includegraphics[width=\linewidth]{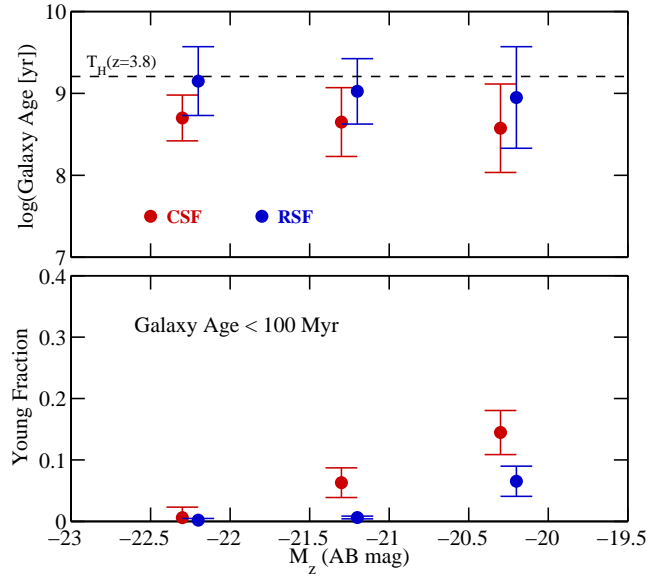}
  \caption{\textit{(Top) } The parameters of the best-fit age distributions as a function of rest-frame optical luminosities. Mean galaxy ages are increasing only by 35\% (60\%) from $M_z=-20.25$ to $M_z=-22.25$, while the width of the age distribution decreases slightly from $\sigma\simeq0.55$ at faint luminosities to $\sigma\simeq0.3-0.4$, depending on the assumed SFH.
  \textit{ (Bottom) } The fraction of galaxies with ages younger than 100 Myr based on the inferred age distributions. Clearly, the brightest galaxies are not dominated by short bursts with $<5\%$ showing galaxy ages younger than 100 Myr. For constant star-formation histories, this increases to 15\% at lower luminosity, while it stays rather small for rising SFHs.
  }
	\label{fig:ageSimSummary}
\end{figure}

\section{Discussion}
\label{sec:discussion}

The large sample of galaxies with reliable IRAC photometry analyzed in this paper allowed us for the first time to study the correlations between UV and optical colors at $z\sim4$, as a function of optical luminosity. The relatively tight correlation we find between the  $J_{125}-[4.5]$ color and the UV slope $\beta$ in Figure \ref{fig:betaJ1} provides direct evidence that the main driver for UV reddening in LBG samples is dust extinction.

The shallow slope of this relation is particularly intriguing, since \citet{Calzetti00} dust reddening would result in significantly steeper trends. This could in principle be the result of correlated changes in galaxy ages with dust extinction, such that younger galaxies are more heavily obscured. However, a more simple explanation is that dust extinction at $z\sim4$ is more accurately described by an SMC-like dust curve. 

Certainly, our measurements are not ideal to determine the exact shape of the reddening law. However, the fact that the brightest galaxies do follow a $J_{125}-[4.5]$ vs $\beta$ trend that is parallel to the mean trend but offset by a constant value strongly suggests that the correlation is driven by SMC dust, and that the offset is due to older stellar population ages compared to the mean.
If this is confirmed in future analyses, the use of SMC-like dust would result in small corrections of the reported SFRs for $z>4$ galaxies in the literature.

While both local Lyman Break Analogs and the majority of bright LBGs at $z\sim2-3$ are more consistent with Calzetti dust extinction \citep[e.g.][]{Overzier11,Reddy11,Reddy10}, SMC-like extinction has been found to better reproduce the infrared excess of young, low-luminosity galaxies at $z\sim2-3$ \citep[e.g.][]{Siana08,Siana09,Reddy11}. On the other hand, direct measurements of the IR excess with Herschel of $z\sim4$ galaxies suggests an inverted luminosity dependence, with brighter galaxies having SMC-like dust, while fainter galaxies are found to follow the M99 relation \citep{Lee12b}. 

Clearly, the appropriate dust curve is still an open question. Solving this problem will require a multi-wavelength approach using complete samples of galaxies with unbiased spectroscopic redshifts, and combining deep $HST$ and $Spitzer$ data with Herschel and ALMA observations. Further progress should thus be possible in the very near future. 

For now, we use the $J_{125}-[4.5]$ vs $\beta$ relation to empirically correct the Balmer break color for dust extinction. While the observed $J_{125}-[4.5]$ colors are significantly redder for the brightest sources, the dust-corrected colors only show a relatively weak dependence on rest-frame optical luminosity. A residual 0.2 mag difference between galaxies at $M_z<-21.75$ and fainter sources is easily explained by a small increase in the stellar population age by only $\sim$0.2 dex. 

Dust extinction is clearly the main contributor to red galaxy colors even at the brightest luminosities, and the steepening of the color-magnitude relation seen in Figure \ref{fig:fig2} is the result of a significant increase in the dust reddening in the most massive galaxies. This could have important consequences for the interpretation of the UV LFs, since increased dust extinction can effectively saturate the UV luminosities \citep[see e.g.][]{Reddy08,Bouwens09b}. 

The steeper color-magnitude relation at $L>L*$ is an indication that the most massive galaxies produce a significantly higher dust mass relative to their stellar mass than sub-$L*$ galaxies, or that they can more easily hold on to the dust produced in their supernovae compared to lower mass systems  \citep[see also][]{Bouwens12a,Finkelstein12}. 
Although \citet{Finkelstein12} only rely on rest-frame UV data to infer stellar masses, they find that the most massive galaxies show surprisingly red UV continuum slopes, which would require significant amounts of dust to be present as early as $z\sim7$. A more robust exploration of this effect is now possible based on the large samples with IRAC photometry (Gonzalez et al., in preparation).

In contrast to the strong trends of dust extinction, the galaxy ages only vary weakly with luminosity.
The inferred mean galaxy ages increase by only 35\% to 60\% across 2 mag in $z$-band luminosity, depending on the assumed SFH (see Figure \ref{fig:ageSimSummary}). This is derived using the same simulations as described in Section \ref{sec:fitstodistribution}.
The best-fit distributions show that there are essentially no bright galaxies with colors consistent with being dominated by very young bursts. However, this fraction appears to be increasing toward fainter luminosities (lower panel Figure \ref{fig:ageSimSummary}), which can further be explored in the near future with the deeper S-CANDELS data.

Unfortunately, current simulations of high-redshift galaxies mostly focus on the sub-L* population, due to the very large volumes necessary to sample the brightest galaxies. However, the very steep increase in dust extinction in $>L*$ galaxies which we find in our data presents an interesting test for galaxy evolution models. For example, we show that the corrected $J_{125}-[4.5]$ color distribution is almost perfectly reproduced by the \citet{Guo11} SAM models, indicating that this SAM is reliably predicting galaxy SFHs. However, these same models fail in accurately capturing the observed steepening in the color magnitude relation and in predicting the UV LF and its evolution. Both these effects are likely connected to the uncertain treatment of dust extinction in such models \citep[see e.g.][]{Gonzalez-Perez12,Henriques12}.

\section{Summary}
\label{sec:summary}

In this paper, we combined ultra-deep HST and Spitzer imaging to study rest-frame UV-to-optical colors of galaxies at $z\sim4$. In particular, we perform a sophisticated neighbor subtraction in the IRAC images to obtain a sample of 1273 B-dropout galaxies with clean IRAC [4.5] detections more significant than 3$\sigma$ from two fields of the HUDF09 survey as well as from GOODS-North and South. 
The [4.5] channel data is preferred over the shorter wavelength [3.6] photometry in order to avoid any flux contamination by strong H$\alpha$ emission \citep[e.g.][]{Shim11,Stark12}.

Using this sample we find:
\begin{itemize}

\item The observed $J_{125}-[4.5]$ color is strongly dependent on rest-frame optical luminosity. In particular, the relation steepens significantly for galaxies with $M_z<-21.5$ compared to the lower luminosity trend (see Figure \ref{fig:fig2}), indicating that the most massive galaxies produce and retain more dust relative to their stellar mass than lower mass galaxies.

\item A relatively tight correlation is found between $J_{125}-[4.5]$ and the UV continuum slope $\beta$ (see Figure \ref{fig:betaJ1}). The most simple explanation for this slope is that the UV-to-optical SEDs of $z\sim4$ galaxies are better reproduced by an SMC-like dust extinction rather than the standard Calzetti dust curve. This result will have to be confirmed in the future by combining large samples of galaxies with spectroscopic redshifts with additional longer-wavelength data to constrain the amount of dust extinction.


\item The dust-corrected colors of galaxies are only weakly dependent on luminosity, and their distribution is well reproduced by a Gaussian in log(age) of galaxies, with a mean log(age)$=8.57$ or 9.00, depending on the assumed star-formation history (either constant or rising with time).

\item Only a very small fraction of $12-19$\% of galaxies shows colors that are consistent with extremely young galaxy ages, below 100 Myr. The exact number of this fraction is dependent on the assumptions about the star-formation histories of galaxies. However, the vast majority of galaxies is consistent with a smooth build-up and only a negligible fraction of their stars being formed in very recent burst.

\end{itemize}

This paper gives a brief preview of the power of combining HST and Spitzer observations for studying galaxies at $z>4$. After completion of the still on-going S-CANDELS program, these types of analyses can be extended to cover more than twice the area of our current sample to a depth of 50 h (about 0.4 mag deeper than the GOODS IRAC data used here). These data sets will allow us to constrain SFH and age distributions of large samples of galaxies at $z>3$. In particular when combined with additional constraints on dust extinction, e.g. from Herschel or ALMA, and with spectroscopic redshifts from the new and upcoming multi-object NIR spectrographs.
This should allow for the next big steps forward over the next few years in working toward a self-consistent picture of star-formation and mass build-up before the peak of cosmic star-formation.

\bigskip

\acknowledgments{Support for this work was provided by NASA through Hubble Fellowship grant HF-51278.01 awarded by the Space Telescope Science Institute, which is operated by the Association of Universities for Research in Astronomy, Inc., for NASA, under contract NAS 5- 26555. This work was additionally supported by NASA grant NAG5-7697 and NASA grant HST-GO-11563.01. }

Facilities: \facility{HST(ACS/WFC3), Spitzer(IRAC)}.

\bibliographystyle{apj}

\appendix

\section{The Effects of Using the Meurer et al. 1999 Dust Correction}

\begin{figure}[tbp]
	\centering
	\includegraphics[width=0.5\linewidth]{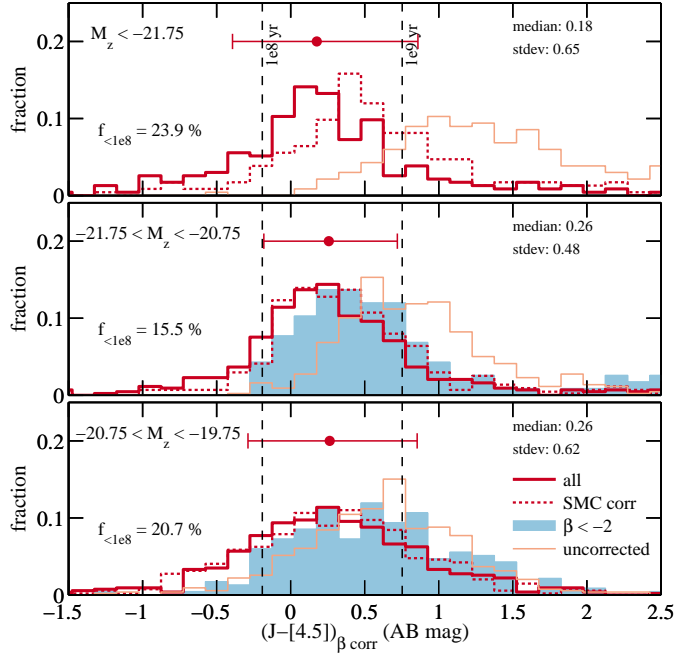}
  \caption{Same as Figure \ref{fig:ColHist}, but showing the dust-corrected colors when using the \citet{Meurer99} $A_{UV}-\beta$ relation and the Calzetti dust curve. For comparison, the dashed line shows the histogram for our fiducial dust correction as described in the main text. As can be seen, using the Calzetti dust correction results in dust corrected colors that are bluest at the brightest luminosities. Furthermore, the corrected colors for the bluest galaxies with $\beta<-2$ are significantly redder than for the full population. If this were true, this would indicate that the bluest galaxies are the oldest in the sample, which is not expected. Therefore, the combination of Meurer and Calzetti dust likely over-corrects the $J_{125}-[4.5]$ colors. Nevertheless, the integrated fraction of galaxies with corrected colors bluer than what is expected for a CSF population at 100 Myr is only 19\%.
  }
	\label{fig:ColHist2}
\end{figure}

As we show in the main text of this paper, an SMC like dust curve appears to be more consistent with the $J_{125}-[4.5]$ vs $\beta$ relation (see Figure \ref{fig:betaJ1}). This motivated our use of an SMC slope to correct galaxy colors for dust. However, for consistency with much of the previous literature, we also show the effects of using the combination of the \citet{Meurer99} relation between $A_{UV}$ and the UV slope $\beta$ together with a \citet{Calzetti00} dust curve. The Meurer corrected color histograms are shown in Figure \ref{fig:ColHist2}.

As can be seen, this approach results in dust corrected colors that are bluest at the brightest luminosities. Furthermore, the corrected colors for the bluest galaxies with $\beta<-2$ are significantly redder than for the full population. If this were true, this would indicate that the bluest galaxies are the oldest in the sample, which is not expected. Therefore, the combination of Meurer and Calzetti dust likely over-corrects the $J_{125}-[4.5]$ colors. 

Despite of the larger correction, the integrated fraction of galaxies with corrected colors bluer than what is expected for a CSF population at 100 Myr is still only 19\%.

\begin{figure}[tbp]
	\centering
	\includegraphics[width=0.5\linewidth]{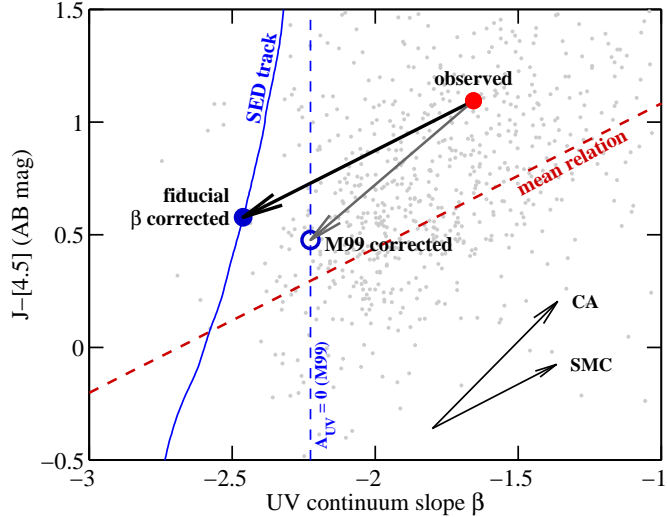}
  \caption{Schematic figure showing our fiducial dust correction procedure. The solid line shows the track of a \citet{Bruzual03} SED model with constant star-formation at $0.2Z_\odot$. For every galaxy we start from the observed location in the $\beta$ vs $J_{125}-[4.5]$ plot (red dot), and we find the intersection of the negative reddening vector with the SED track (filled blue dot) to determine the dust-corrected $J_{125}-[4.5]$ color. For comparison, the standard \citet{Meurer99} correction together with a \citet{Calzetti00} dust curve is also shown by the dark gray vector, which would result in the open blue dot. In particular for very UV-red galaxies the difference between the two dust corrections can be significant. The M99 relation assumes a dust-free UV slope of $\beta=-2.23$ (vertical dashed line), which is substantially redder than the values measured for star-forming SEDs with ages $\lesssim1$Gyr. For the few galaxies that lie to the left of the SED track, we do not apply any dust correction to the color. The red dashed line represents the mean relation between $\beta$ and $J_{125}-[4.5]$ of Equation \ref{eq:1}, shown in Figure \ref{fig:betaJ1}.
  }
	\label{fig:schematicDustCorr}
\end{figure}

\end{document}